\documentstyle[12pt]{article}
\addtocounter{equation}{0}
\begin{document}
\begin{center}
{\bf A Longshot Technique for Resurrecting the Dead by Reversing the\\ Flow of
Time}\\
\hspace*{.2in}\\
H. Hickman\\
\hspace*{.2in}\\
{\em Science Department, Hillsborough Community College, P.O. Box 30030,\\
Tampa, FL \hspace*{.12in}33630-3030}\\
\hspace*{.2in}\\
{\bf Abstract}
\end{center}
Reversing the flow of time between Casimir plates raises the question of
whether or not a recently deceased, intact organism could be brought back to
life. The odds are not good.
\begin{center}
{\bf Discussion}
\end{center}

According to special relativity, a ``stationary'' observer would determine
the energy content of a mass moving with constant relative velocity to be
\begin{equation}
E = mc^2 = \left(\frac{m_o}{\sqrt{1 - \frac{v^2}{c^2}}}\right)c^2\;.
\end{equation}
That same observer would also find
\begin{equation}
\Delta{t} = \Delta{t_o}\sqrt{1 - \frac{v^2}{c^2}}\;,
\end{equation}
where $\Delta{t}$ represents the number of seconds that goes by on the moving
mass, and $\Delta{t_o}$ is the number of seconds that goes by in the stationary
frame.\\

Multiplying equation (2) by equation (1) results in
\begin{equation}
E\Delta{t} = \left(m_oc^2\right)\Delta{t_o} = E_o\Delta{t_o}\;.
\end{equation}
Evidently the energy time-flow product is invariant under special relativity
such that
\begin{equation}
E_1\Delta{t_1} = E_2\Delta{t_2}
\end{equation}
from frame to frame.\\

Equation (4) also applies to a photon leaving a Schwarzschild metric, and
two possibilities come to mind. Either equation (4) is universally applicable
(i.e.- beyond the confines of special relativity), or there must be two
different kinds of energy; the kind that is accompanied by time dilation,
and the kind that is not accompanied by time dilation.\\

If equation (4) does apply everywhere, and if the energy density between
Casimir plates is truly negative, then it may be possible to reverse the
flow of time between Casimir plates in accordance with$^1$
\begin{equation}
E_1\Delta{t_1} = \left(-E_2\right)\left(-\Delta{t_2}\right)\;.
\end{equation}

That possibility brings to mind an interesting question. Suppose some tiny
biological organism - small enough to fit between the two plates - was
allowed to expire from ``old age''. Could that organism then be brought
back to life by immediately placing it in the region of negative time flow
between the plates ?.\\

Several considerations argue against this happening. For one thing, the energy
density of the dead organism would still be positive, even though the organism
was immersed in a region of negative energy density. So time might still flow
forwards inside the organism. On the other hand, the time direction probably
cannot contain jump discontinuities. There would have to be some sort of
smooth transition at the surface of the organism between the region of forward
time flow inside the organism, and the region of backward time flow outside
the organism. So the overall efect is not clear. Further, even if time did
flow backwards throughout the entire organism, there is no guarantee that an
``instant replay'' effect would occur. The components of the dead organism
might just get younger without ``rewinding the tape'' and rekindling the spark
of life.
\begin{center}
{\bf References}
\end{center}
$^1$ H. Hickman, ``Reversing the Flow of Time Between Casimir Plates'',
{\em Speculations in Science and Technology} {\bf 19}(4), 281 (1996). The
original paper has the directions of plate orientation backwards. That
mistake was addressed in a {\em Corrigendum} published later on.

\end{document}